# On Randomness of Goldbach Sequences

Krishnama Raju Kanchu and Subhash Kak

*Abstract-* We consider the use of Goldbach numbers as random sequences. The randomness is analyzed in terms of the autocorrelation function of the sequence of number of partitions. The distinct representations of an even number *n* as the sum of two primes is a local maximum for multiples of the product of the consecutive smallest primes less than the number. Specific partitions, which we call Goldbach ellipses, are examined. It is shown that such ellipse sequences also have excellent randomness property.

## I. INTRODUCTION

Number-theoretic functions provide excellent random sequences that are invaluable in cryptography. Such random sequences may be used for generation of keys and as pseudorandom sequences. The use of modular exponentiation and elliptic curves [1] and even more elementary functions [2],[3],[4] to generate random sequences is well-established. Recently, primitive Pythagorean triples were proposed as random numbers [5]. Other number theoretic sequences may be similarly used.

Christian Goldbach, in a letter to Euler in 1742, suggested that every integer greater than 6 is the sum of three primes. Euler responded that since every even number 2n+2 must have 2 as one of these primes, therefore every even number is a sum of two primes. Goldbach's idea is now stated in two different ways. According to the even Goldbach conjecture, any even number greater than 2 can be represented as sum of two primes. The odd Goldbach conjecture states that any odd number greater than 5 can be expressed as sum of three primes. The falsity of the Goldbach conjecture requires that for *n,* all instances of $n-p_i$, for consecutive odd primes $p_i$ up to $n/2$ be non-prime. According to the prime number theorem, an integer *m* selected at random has roughly 1/ln *m* chances of being prime or (1-1/ln *m*) of being non-prime. The probability that the conjecture is false is therefore $\prod_i (1-\frac{1}{p_i})$ which becomes progressively smaller as n increases. The conjectures have been numerically validated to be true up to $2 \times 10^{17}$ and widely believed to be generally true.

Here we are concerned only with representation of even numbers by prime partitions. Examples of prime partitions of even numbers are:

    4 = 2+2
    6 = 3+3
    8 = 3+5
    10 = 3+7 or 5+5

......

The ways a number can be partitioned generally increases with the number. Additive partitioning has parallels with prime factors for which the earliest reference goes back to ancient Indian texts [6]. Euclid gave a clear definition of a prime number in his Elements (book 7, definition 11).

Let *g(n)* be the number of unique ways n can be expressed as *p+q*, where *p* and *q* are primes. As the value of the number *n* increases, the number of pairs *g(n)* that produces the sum also increases. Table 1 presents some values of *g(n)*:

**Table 1.** Number of partitions *g(n)* of number *n*

| n | 4 | 6 | 8 | 10 | 12 | 14 | 16 | 18 | 20 | 22 | 24 | 26 | 28 | 30 | 32 | 34 | 36 | 38 | 40 | 42 | 44 | 46 | 48 | 50 | 52 | 54 | 56 |
|---|---|---|---|---|---|---|---|---|---|---|---|---|---|---|---|---|---|---|---|---|---|---|---|---|---|---|---|
| g(n) | 1 | 1 | 1 | 2 | 1 | 2 | 2 | 2 | 2 | 3 | 3 | 3 | 2 | 3 | 2 | 4 | 4 | 2 | 3 | 4 | 3 | 4 | 5 | 4 | 3 | 5 | 3 |

The sequence *g(n)* may be converted in to a binary sequence *b(n)* by mapping each even number to 0 and each odd number to 1.

This paper presents some properties of *g(n)* that are of relevance from the perspective of randomness and cryptography. It is shown that *b(n)* sequences have excellent autocorrelation properties. We show that the distinct representations of an even number n as the sum of two primes is a local maximum for multiples of the product of the consecutive smallest primes less than the number. Some properties of Goldbach ellipses are described.

## II. NUMBER OF PARTITIONS

The theory of partitions provides estimates on the function *g(n)*. Figures 1 and 2 give examples of the function to the values 500 and 1,000. The function grows in a band.

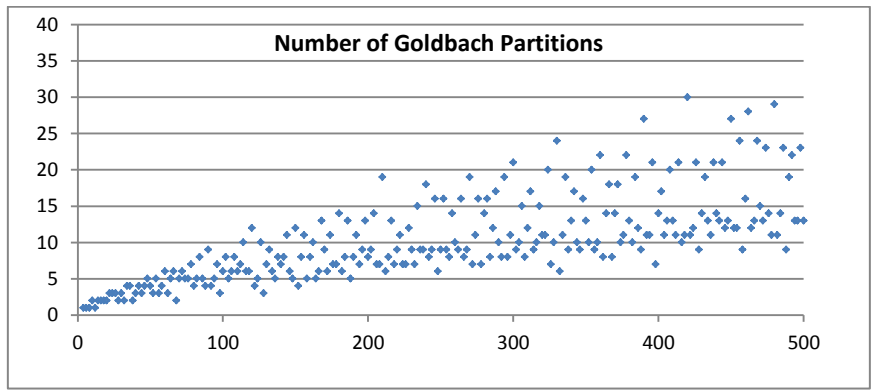

**Figure 1.** Number of Goldbach partitions g(n) for n ≤ 500

A formula estimating *g(n)* may be derived using statistical considerations. For large *n* and *m* between *3* and *n/2*, the probability of *m* and *n-m* being prime is given by 1/[ln *m* ln*(n-m)*]. Thus, the number of ways a large even number *n* can be expressed as a sum of two odd primes is approximately equal to

$$\sum_{m=3}^{n/2} \frac{1}{\ln m \ln(n-m)} \approx \frac{n}{2\ln^2 n}$$

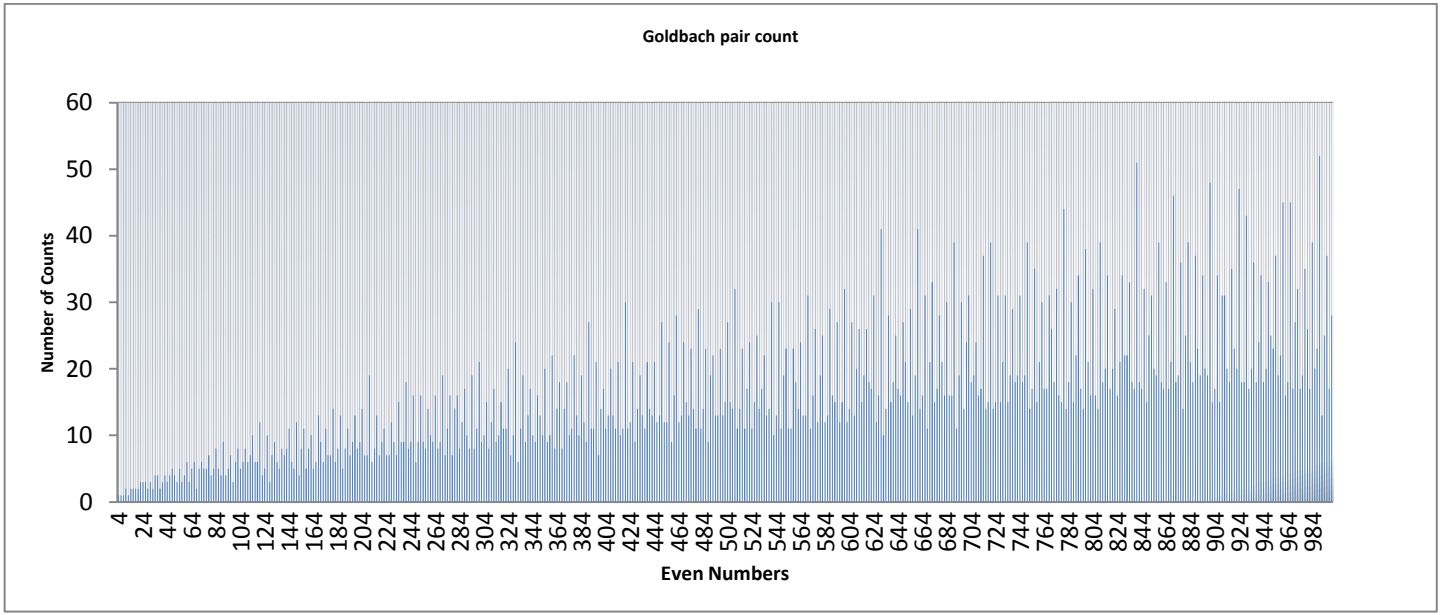

**Figure 2**. Number of Goldbach partitions g(n) for n ≤ 1000

But this expression may, in general, not include dependencies amongst numbers. For example, for an even number *n* that is divisible by 3, and *m* prime, *n-m* is coprime to 3 and thus more likely to be prime rather than a general number. If corrections are used, we obtain the estimate [7]:

$$2\pi_2 \left( \prod_{p|n; p\geq 3} \frac{p-1}{p-2} \right) \frac{n}{\ln^2 n}$$

where $\pi_2$ is the twin prime constant $\prod_{p\geq 3}\left(1-\frac{1}{(p-1)^2}\right) = 0.6601618158...$

Table 2 presents the number of Goldbach partitions for numbers up to 188. Over this range the number varies between 1 and 14. For numbers less than 2000, the largest number of partitions is 91 for n=1890.

**Table 2.** Number of partitions up to n= 188

| n | f(n) | n | f(n) | n | f(n) |
|---|------|-----|------|-----|------|
| 4 | 1 | 66 | 6 | 128 | 3 |
| 6 | 1 | 68 | 2 | 130 | 7 |
| 8 | 1 | 70 | 5 | 132 | 9 |
| 10 | 2 | 72 | 6 | 134 | 6 |
| 12 | 1 | 74 | 5 | 136 | 5 |
| 14 | 2 | 76 | 5 | 138 | 8 |
| 16 | 2 | 78 | 7 | 140 | 7 |
| 18 | 2 | 80 | 4 | 142 | 8 |
| 20 | 2 | 82 | 5 | 144 | 11 |
| 22 | 3 | 84 | 8 | 146 | 6 |
| 24 | 3 | 86 | 5 | 148 | 5 |
| 26 | 3 | 88 | 4 | 150 | 12 |
| 28 | 2 | 90 | 9 | 152 | 4 |
| 30 | 3 | 92 | 4 | 154 | 8 |
| 32 | 2 | 94 | 5 | 156 | 11 |
| 34 | 4 | 96 | 7 | 158 | 5 |
| 36 | 4 | 98 | 3 | 160 | 8 |
| 38 | 2 | 100 | 6 | 162 | 10 |
| 40 | 3 | 102 | 8 | 164 | 5 |
| 42 | 4 | 104 | 5 | 166 | 6 |
| 44 | 3 | 106 | 6 | 168 | 13 |
| 46 | 4 | 108 | 8 | 170 | 9 |
| 48 | 5 | 110 | 6 | 172 | 6 |
| 50 | 4 | 112 | 7 | 174 | 11 |
| 52 | 3 | 114 | 10 | 176 | 7 |
| 54 | 5 | 116 | 6 | 178 | 7 |
| 56 | 3 | 118 | 6 | 180 | 14 |
| 58 | 4 | 120 | 12 | 182 | 6 |
| 60 | 6 | 122 | 4 | 184 | 8 |
| 62 | 3 | 124 | 5 | 186 | 13 |
| 64 | 5 | 126 | 10 | 188 | 5 |

## III. AUTOCORRELATION FUNCTION

The autocorrelation function is a measure of the randomness associated with a time-series. In general it gives the correlation for the signal at two different points.

$$C(s,t) = corr\ (X(s), X(t))$$

If $C(s,t)$ is two-valued, that is large for $s=t$ and close to zero for $s \neq t$, then it may be taken to be random. This is in analogy with a random process.

For a stationary process, only the difference between the two points needs to be considered and one may simplify the function for a discrete sequence $a_j$, $j=0,1,...n-1$, to:

$$C(k) = 1/n \sum_{j=0}^{n-1} a_j a_{j+k}$$

A deterministic sequence that satisfies near-two-valued property is considered pseudorandom. The autocorrelation function of the Goldbach sequence (converted in to a binary sequence by considering even numbers as -1 and odd numbers as 1) for n=2000, is given in Figure 3. The autocorrelation function has values for non-zero $k$ that is much less than 10% of the peak value at $k=0$.

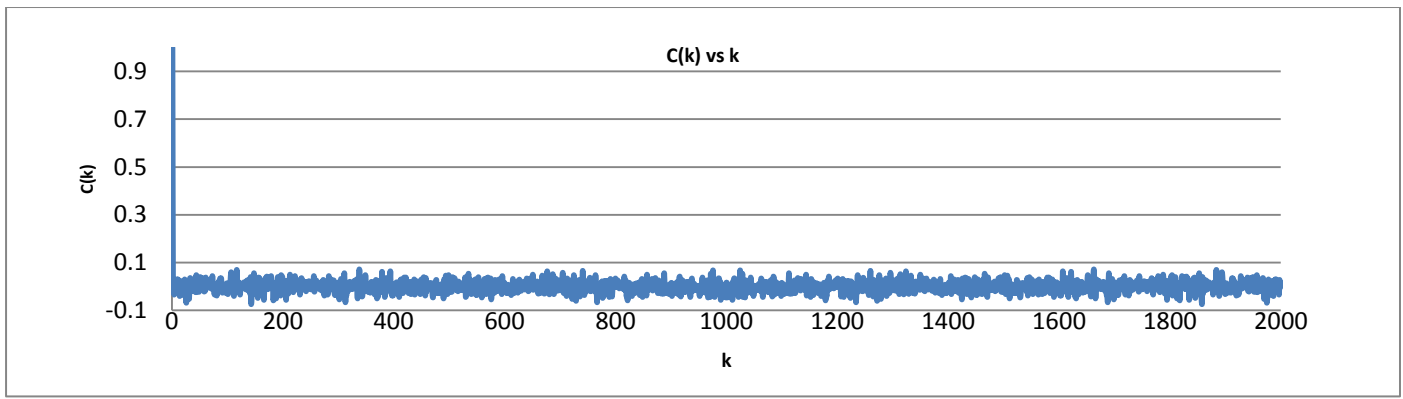

**Figure 3**. Autocorrelation function for Goldbach sequence when n=2000

## IV. PARTITIONS BASED ON THE DIFFERENCE OF TWO PRIMES

The standard Goldbach's conjecture considers sums of two primes. Now we generate random numbers based on the Goldbach pairs that would be produced as a result of difference of two primes. The number *n* is represented as:

$$n = p_1 - p_2$$

In the example given in Figure 4, we consider the last prime number to be 2003 and we find the prime pairs for even numbers over the range 4 to 2002. The number of primes with a difference of 2 is 35, with a difference of 4 is 65, and with a difference of 6 is 129. Figure 4 is roughly the mirror image of Figure 2.

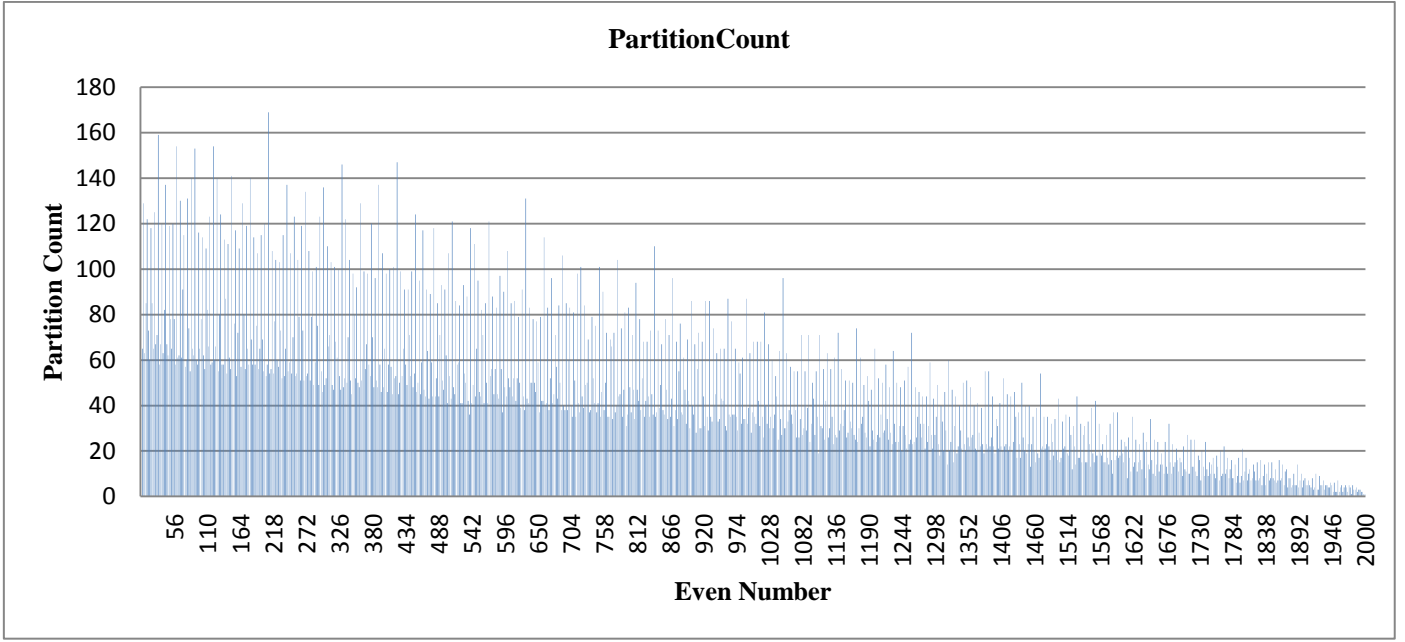

**Figure 4.** Count of number of representations for numbers represented as difference of primes

The peaks of Figure 4 have cycles identical to that of Figure 2.

## V. FURTHER PROPERTIES OF THE PARTITION FUNCTION

The partition function $g(n)$ has a local peak when the number of factors of $n$ is large. We will thus get peaks at multiples of 2×3=6; 2×3×5=30; 2×3×5×7=210; 2×3×5×11=330; 2×3×5×13=390; 2×3×5×7×11=2310; 2×3×5×7×11×13=30030; 2×3×5×7×11×13×17=510510; 2×3×5×7×11×13×17×19=9699690, etc.

Table 3 illustrates the difference for multiples of these numbers and their adjacent values. The difference between the peak and the next values ranges between three and four times.

Due to these peaks, we get other conditions such as:

$g(6k) > g(6k+2)$

$g(30k) > g(30k+2)$

$g(210k) > g(210k+2)$

and so on…

**Table 3.** The number of partitions near 30030, 60060, 90090, and 1021020

| n | g(n) | n | g(n) | n | g(n) | n | g(n) |
|---|---|---|---|---|---|---|---|
| 30020 | 318 | 60050 | 524 | 90080 | 741 | 1021010 | 5567 |
| 30022 | 240 | 60052 | 397 | 90082 | 577 | 1021012 | 4163 |
| 30024 | 470 | 60054 | 798 | 90084 | 1119 | 1021014 | 8402 |
| 30026 | 223 | 60056 | 406 | 90086 | 578 | 1021016 | 4518 |
| 30028 | 237 | 60058 | 410 | 90088 | 552 | 1021018 | 4127 |
| 30030 | 905 | 60060 | 1564 | 90090 | 2135 | 1021020 | 17075 |
| 30032 | 225 | 60062 | 387 | 90092 | 552 | 1021022 | 4401 |
| 30034 | 224 | 60064 | 394 | 90094 | 547 | 1021024 | 4140 |
| 30036 | 466 | 60066 | 846 | 90096 | 1110 | 1021026 | 8228 |
| 30038 | 232 | 60068 | 400 | 90098 | 594 | 1021028 | 4179 |

For *n*=6, 30, and 210, all primes in the range [*n/2, n-2*] are amongst the partitions. The number n=210 is the largest such number possible [8].

The Goldbach partitions of 30 are (23,7), (19, 11), (17, 13) and that of 210 are

(199, 11), (197, 13), (193, 17), (191, 19), (181, 29), (179, 31), (173, 37), (167, 43), (163, 47), (157, 54), (151, 59), (149, 61), (139, 71), (137, 73), (131, 79), (127, 83), (113, 97), (109, 101), (107, 103)

The Goldbach partitions of 420 cover all primes in the range [*n/2, n-2*] excepting 277, 251, and 233. For the case of n=630, all primes in the corresponding range are covered excepting these 8 cases: 509, 487, 461, 443, 421, 409, 383, and 331.

## VI. GOLDBACH ELLIPSE

Let *2n = p+q*, as sum of two primes, where *n-p=q-n*. The Goldbach radius of the number *n* is then the smallest *n-p* associated with the number. As example, consider n=14. The number 28 can be expressed in two ways as sum of primes that are equidistant from 14: 5+23 and 11+17. The Goldbach radius of 14 is then the smaller of the two sets, which is 3. This radius, therefore, picks one of the Goldbach partitions associated with the number and it associates a circle with the number.

A generalization of the circle is the ellipse where the distances from the components on two sides of the even number are different. For simplicity, we consider the ellipse *(1,k)*. It represents the smallest neighbors of the given number *n* that is co-prime with *k* that are a prime partition of the given number where the distances are multiples of 1 and *k*, respectively. The characteristic numbers are: *2n, 2n-m, 2n+km, 4n+(k-1)m.*

When k=1, we have the case of the circle since the partitions add up to 2*n*.

The mapping of the sequence *2n* into m or *4n+(k-1)m* can have cryptographic applications. By choosing a subset of *2n*, one can generate a corresponding set of numbers *m*. From a sufficiently small subsequence of such numbers, the original sequence cannot be determined.

**Example 1.** Consider *k=7*. The partitions of numbers n that are not multiples of 7 are as follows:

**Table 4.** Partitions in an ellipse for *k=7*

| 2n | 2n-m | 2n+km | m | 4n+(k-1)m |
|----|------|-------|---|-----------|
| 4  | 3    | 11    | 1 | 14        |
| 6  | 5    | 13    | 1 | 18        |
| 8  | 5    | 29    | 3 | 34        |
| 10 | 7    | 31    | 3 | 38        |
| 12 | 11   | 19    | 1 | 30        |
| 16 | 13   | 37    | 3 | 50        |
| 18 | 13   | 53    | 5 | 66        |
| 20 | 17   | 41    | 3 | 58        |
| 22 | 19   | 43    | 3 | 62        |
| 24 | 23   | 31    | 1 | 54        |
| 26 | 23   | 47    | 3 | 70        |
| 30 | 29   | 37    | 1 | 66        |
| 32 | 29   | 53    | 3 | 82        |
| 34 | 19   | 139   | 15| 158       |

For numbers that are multiples of *k*, no ellipse is defined as *2n+km* will not be prime. This is the reason that numbers that are multiples of 7 are not in the left-most column.

**Example 2.** Consider k=3. The partitions of numbers n that are not multiples of 3 are as follows:

Table 5. Partitions in an ellipse for k=3

| 2n | 2n-m | 2n+km | m | 4n+(k-1)m |
|----|------|-------|---|-----------|
| 4  | 3    | 7     | 1 | 10        |
| 8  | 7    | 11    | 1 | 18        |
| 10 | 7    | 19    | 3 | 26        |
| 14 | 13   | 17    | 1 | 30        |
| 16 | 11   | 31    | 5 | 42        |
| 20 | 19   | 23    | 1 | 42        |
| 22 | 19   | 31    | 3 | 50        |
| 26 | 19   | 47    | 7 | 66        |
| 30 | 19   | 43    | 3 | 62        |
| 32 | 29   | 41    | 3 | 70        |
| 34 | 31   | 43    | 3 | 74        |

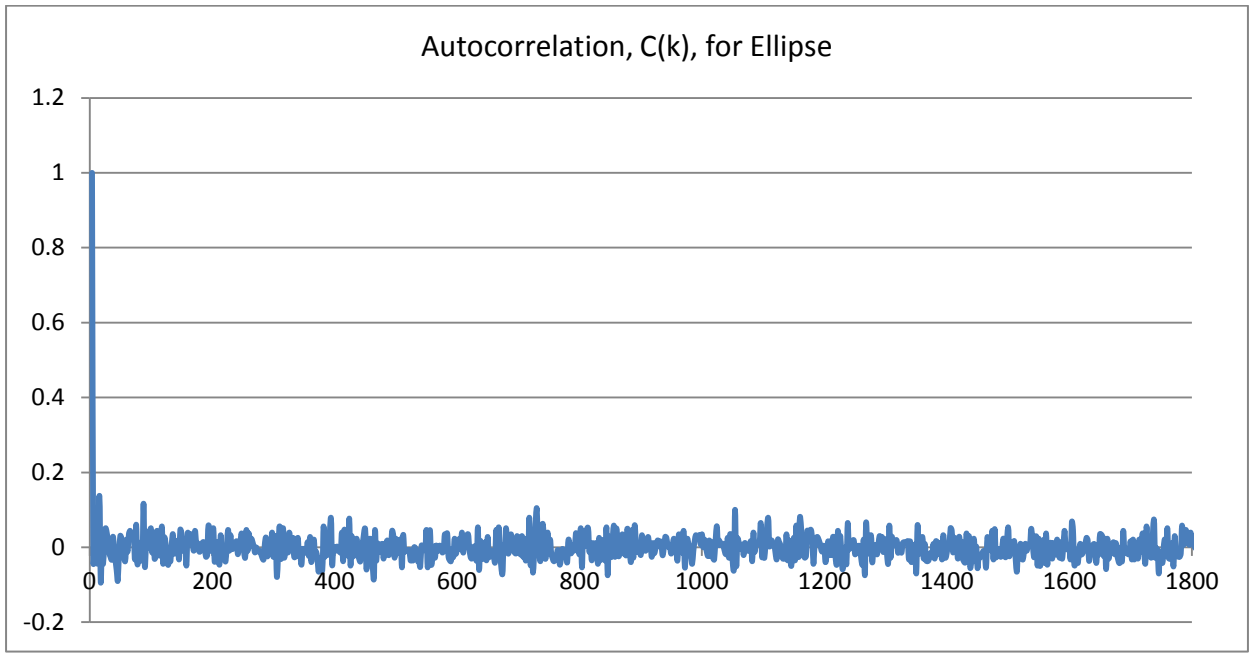

**Figure 5.** Autocorrelation function for binary sequence generated from Table 4 that has been mapped with respect to 2n

Figure 5 presents the autocorrelation function of the sequence in the fourth column of Table 4 with numbers m. The numbers in this column are reduced to 1 and -1 based on whether they are 1 mod 4 or 3 mod 4. Thus the first few numbers of the sequence in Table 4 are 1 1 -1 -1 1 -1 1 -1 -1 1 -1 1 -1 -1 and so on.

## VII. CONCLUSION

We have reviewed some properties of Goldbach sequences that are of relevance in randomness. We have presented a procedure for generating a random sequence using Goldbach partitions. Goldbach ellipses and circles associated with even numbers are described. We have shown that the Goldbach sequence as well as the sequence obtained from the Goldbach ellipse have good autocorrelation properties and, therefore, they can be used in cryptographic applications.